\documentstyle[referee,epsfig,graphicx]{mn2e}
\newif\ifAMStwofonts
\newcommand{\be}{\begin{equation}}
\newcommand{\ee}{\end{equation}}
\newcommand{\bea}{\begin{eqnarray}}
\newcommand{\eea}{\end{eqnarray}}

\newcommand{\da}{\frac{\delta}{\alpha}}
\newcommand{\di}{\partial_}
\newcommand{\ad}{\frac{\alpha}{\delta}}

\title{Coarse-graining the distribution function of cold dark matter II }

\author[ R.N. Henriksen]
{
R.N. Henriksen$^1$\thanks{henriksn@astro,queensu.ca}
\\
$^1$Queen's University, Kingston, Ontario, K7L 3N6,Canada \\
and SAp/DAPNIA, CEA Saclay, 91191 Gif-sur Yvette, CEDEX, France\\
}
\date{\today}
\pagerange{\pageref{firstpage}--\pageref{lastpage}}
\pubyear{2004}

\begin{document}

\maketitle

\label{firstpage}

\begin{abstract}
   
 We study analytically the coarse and fine-grained distribution function established by the self-similar infall of collisionless matter. We find this function explicitly for isotropic and spherically symmetric systems in terms of cosmological initial conditions. The coarse-grained function is structureless and steady but the familiar phase space sheet sub-structure is recovered in the fine-grained limit. By breaking the self-similarity of the halo infall we are able to argue for a central density flattening. In addition there will be an edge steepening. The best fitting analytic density function is likely to be provided by a high order polytrope fit smoothly to an outer power law of index  $-3$ for isolated systems. There may be a transition to a $-4$ power law in the outer regions of tidally truncated systems.

 Since we find that the central flattening is progressive in time, dynamically young systems such as galaxy clusters may well possess the NFW type density profile, while primordial dwarf galaxies for example are  expected to have cores. 
This progressive flattening is expected to end either in the non-singular isothermal sphere, or in the non-singular metastable polytropic cores; since the distribution functions associated with each of these arise naturally in the bulk halo during the infall. We suggest  based on previous studies of the evolution of de-stabilized polytropes, that a collisionless system may pass through a family of polytropes of increasing order, finally approaching the limit of the non-singular isothermal sphere, if the `violent' collective relaxation  is frequently re-excited by  `merger' events. Thus cD galaxies and indeed all bright galaxies that have grown in this fashion, should be in polytropic states.  
Our results suggest that no physics beyond that of wave-particle scattering is necessary to explain the nature of dark matter density profiles. However this may be assisted by the scattering of particles from the centre of the system by the infall of dwarf galaxies, galactic nuclei or black holes (e.g. Nakano and Makino, 1999), all of which would restart the pure dynamical relaxation. 
     
\end{abstract} 
\begin{keywords}
methods: analytical--gravitation--accretion--distribution functions.
\end{keywords}
\setlength{\baselineskip}{13pt}
\section{Introduction}
\label{sec:intro}
In a previous paper (Henriksen and Le Delliou, 2002 $\equiv$ paper I) a novel analytic method for studying the (mass) Distribution Function (DF) of a collisionless self-gravitating system of particles was introduced. The essential idea is to carry out an expansion of the coupled Boltzmann-Poisson system in terms of the phase space resolution. Normally the zeroth order in the expansion corresponds to the coarsest-grained view, while the higher order terms correspond to less crude coarse-graining. The variable resolution of the system phase space is achieved by means of appropriate non-canonical coordinate transformations, these being chosen to contain a parameter whose magnitude adjusts the minimum phase space volume resolved.  A fully relaxed system can be expected to look the same at different levels of coarse-graining  and so this method permits the construction of equilibrium distribution functions by determining the conditions under which all structure at any level of coarse-graining vanishes. A fine-grained expansion also exists wherein an expansion about the coarse-grained structureless DF in a series of increasing resolution recovers the phase-space structure known to exist in the infall models. 

It should be noted that the value $1$ of the expansion parameter $\alpha$ corresponds to the phase space volume that is used to define a smooth DF. For example, were we applying the technique to a clumped dark matter system or to a system of globular clusters, there would be many individual objects in the phase space volume used in the definition. Larger volumes correspond to an even coarser graining and so terminating the series in $1/\alpha$ at zeroth order is a means of confirming the smoothed nature of the original definition. We obtain fine structure by expanding in values of $\alpha<1$, as is shown below. This provides assurance that the resolution expansion can recover essential features of the DF. 
 
One knows however that the strict self-similarity in the infall must be broken near the centre of the system (paper I and below, where our series may be seen not to converge as the potential gradient diverges). This leads us to study the explicit time dependence of the system as the direct means of breaking the self-similarity (Carter and Henriksen,1991) and this predicts a central flattening in time. In Merrall and Henriksen (2003) a de-stabilized polytrope was found to relax towards a gaussian DF with dispersion that increased in time. That is the `temperature' of the system increased until a new equilibrium was achieved. This effect is also observed in our analytic study of the central relaxation.

The technique for generating the structureless DF is best employed for steady state systems since one expects relaxed systems to be steady. However there is a particular case of time dependence that is of importance since it corresponds to a possible growth mode of dark matter halos. This is the well studied self-similar infall  (e.g. Henriksen and Widrow, 1999 and references therein, Le Delliou and Henriksen, 2003), which is known to arise inevitably for radial orbits and can be assumed as a general model. In the appropriate variables this growing mode of the system can be treated as steady, and in fact we find that the DF's constructed in this way are absolutely steady (indeed they may be constructed by this double requirement; Henriksen and Widrow, 1999). Thus they represent equilibrium states of the system, at least until they are subsequently disturbed. 

A first study of these systems was conducted in paper I. In particular that paper applied the method to  dark matter halos in spherical symmetry to show that the self-similar density profile had to flatten at the centre of the system. It was found that the flattening would be progressive, with the effective power law of the density profile passing through the NFW (Navarro, Frenk and White, 1996) value of $-1$ on the way to even flatter values. This conclusion seems now to be in agreement with the numerical simulations (e.g. Power et al., 2003), which fact offers {\it  confirmation  to both methods}.

The discussion of isotropically elliptical orbits in paper I is unfortunately flawed by the omission of a factor $\zeta_1$ in the last term of equation (80) of that paper that was carried through to the end of that section. Since this term vanishes when $\ad=3$, the discussion of that case remains valid. The general formulae are corrected in the appendix to this paper.

In the bulk of the present paper we present a conclusive application of our ideas to an isotropic, spherically symmetric collisionless dynamical system. This case is likely to be a general ultimate equilibrium since theory, as elaborated e.g. by Lynden-Bell (1967), Aly (1989), Nakamura (2000), suggests that the ultimate relaxed state of a collisionless system should be spherically symmetric with an isotropic Gaussian DF in energy. However our techniques may be applied to a general system and this is discussed in a mathematical appendix. This allows a discussion of axial symmetry and of the dependence on phase-space dimensionality in general, but the physical conclusions are not different from those presented in the text when considerations of regularity, symmetry and entropy are applied.

 The mass distribution of the dominant dark matter in many dwarf elliptical galaxies is best fit by that of the non-singular isothermal sphere (e.g. Cot\'e, Carignan \& Freeman, 2000; Begum, Chengalur and Hopp, 2003; Begum and Chengulur, 2004). 
On the other hand XMM and Chandra studies of hydrostatic hot gas in galaxy clusters (e.g. Buote, 2004; Pratt and Arnaud, 2003, 2002) show in most cases that the NFW profile gives a good fit to the total mass distribution down to about 
$.03$-$.05$ $ r_{200}$ or about $40$- $70$ kpc for the low mass cluster Abel 1983. However the fact that the virial radius $r_{200}$ is only about 4 times the core radius suggests that perhaps the `real' core has not been detected. Based on the results of the present paper we would expect such a low mass cluster to have a core relatively large compared to the high mass clusters. 

Sand et al. (2002) and Gavazzi et al. (2003) have however studied the cluster 
MS 2137.3 -2353 using gravitational lensing of background sources and conclude that the central power law is as flat as $-0.35$. This has been `explained' in terms of infalling baryonic clumps (El-Zant et al. 2004) but we find that such flattening occurs naturally during the relaxation of the dark matter infall and that there is no need for such extrnal influences. It is undoubtedly the case that such `clumpy' infall or merging does have a flattening influence (e.g. Nakano and Makino, 1999), and we note that this happens automatically in the self-similar infall since shells that fall in later have more mass (for $\ad<3/2$).  

A subsequent paper will attempt to convert our predictions as a function of dynamical age to a mass dependence. In the present work we wish principally to justify the method and the formal results.
In the next section we present our results in the context of an isotropic, spherically symmetric halo, since these correspond to our general results and are more accessible. The appendix will provide more mathematical detail for the general case.
 
\renewcommand{\textfraction}{0}
\renewcommand{\topfraction}{1} 
\renewcommand{\bottomfraction}{1}

\section{Isotropic, Spherically Symmetric, Halos} \label{sec:1D}

We present  in this section the formulae that are common to both the general case treated in the appendix and to the spherically-symmetric isotropic case treated here. 

\subsection{Mathematical formulation}\label{sec:geneqs}

As usual we employ a mass distribution function  and here it is taken to be in six phase space dimensions so that $f=f(\vec r,\vec v;t)$. Thus we have to solve
\be
\partial_t f+\vec v.\vec\nabla f+\frac{d\vec v}{dt} . \vec\nabla_v f=0,
\ee
and 
\be
\nabla^2\Phi=4\pi G \rho,
\ee
where 
\be
\rho=\int~f~d^3v.
\ee
 For some applications it is possible to add an external potential to the `internal' $\Phi$. This could be due to a central point mass or to a distribution of baryons.

In spherical polar coordinates  we must remember that 
\be
\frac{d\vec v}{dt}=(\frac{v_\theta^2+v_\phi^2}{r}-\partial_r\Phi,\frac{v_\phi^2\cot(\theta)-v_rv_\theta-\partial_\theta\Phi}{r},-\frac{v_\phi(v_r+v_\theta\cot(\theta))+\frac{1}{\sin(\theta)}\partial_\phi\Phi}{r}).\label{acceleration}
\ee

The variable transformations that render the problem stationary during the self-similar infall are:
\bea
 R= r e^{-\delta T},~~\vec Y=&\vec v e^{-(\delta-\alpha)T},~~ \Psi&=\Phi e^{2(\delta-\alpha)T},\nonumber \\
P = f e^{(3\da-1)\alpha T},~~ \Theta=&\rho e^{(2\alpha )T},& e^{\alpha T}=\alpha t .\label{transsphers}
\eea
The angles $\theta$ and $\phi$, being dimensionless, do not transform.

The equations in these transformed variables become
\bea
t\partial_t P &-&(3\da-1)P+(\frac{\vec Y}{\alpha}-\da\vec R).\vec\nabla P+\left(\frac{Y_\theta^2+Y_\phi^2}{\alpha R}-(\da-1)Y_R-\frac{\partial_R\Psi}{\alpha}\right)\partial_{Y_R}P\nonumber\\
&+& \left(\frac{Y_\phi^2\cot(\theta)-Y_RY_\theta-\partial_\theta\Psi}{\alpha R}-(\da-1)Y_\theta\right)\partial_{Y_\theta}P\nonumber\\
&-&\left(\frac{Y_\phi(Y_R+Y_\theta\cot(\theta))+\frac{1}{\sin(\theta)}\partial_\phi\Psi}{\alpha R}+(\da-1)Y_\phi\right)\partial_{Y_\phi}P=0,\label{genVlasov}
\eea
and
\be
\nabla_R^2\Psi=4\pi G~\Theta,\label{Poisson}
\ee
where
\be
\Theta=\int~P~d^3Y,
\ee
and the $R$ indicates the scaled radial coordinate (the angular coordinates are unchanged).

For the isotropic and spherically symmetric limit that we study in this paper, we set 
\be
P=P(Y^2,R;t)~~~~~~\Psi=\Psi(R;t),
\ee
in the previous equations, where 
\be
Y^2=Y_R^2+Y_\perp^2=Y_R^2+Y_\theta^2+Y_\phi^2.
\ee
Then from a direct substitution into equation (\ref{genVlasov}) we obtain 
\be
t\partial_t P-(3\da-1)P+(\frac{Y_R}{\alpha}-\da R)\partial_R P-(\da-1)Y\partial_Y P-(\frac{1}{\alpha}\partial_R\Psi)(\frac{Y_R}{Y})\partial_Y P=0,
\label{isospherVlasov}
\ee
and of course equation (\ref{Poisson}) becomes simply
\be
 \frac{1}{R^2}\frac{d}{dR}\left(R^2\frac{d~\Psi}{d~R}\right)=4\pi G~\Theta.
\label{spherPoisson}
\ee

The coarse-graining or fine-graining is carried out in terms of the parameter $\alpha$ since it is easy to show from the transformations of equation (\ref{transsphers}) that the physical phase space element is proportional to the factor $\exp{(6\da-3)\alpha T}$ multiplying the phase space element in transformed variables. Thus for a fixed self-similar index ($\da$) we may increase or decrease the physical phase space volume (relative to the reference value at $\alpha=1$) corresponding to a given set of transformed variables by increasing or decreasing $\alpha$ respectively. This is true for $\ad<2$. This limiting value $2$ renders the transformation canonical so that no transformation of the phase space volume is effected. We can nevertheless approach this limit arbitrarily closely; but in any case, as discussed in the appendix, an expansion in powers of $1/\alpha$ amounts to an expansion in time resolution and therefore the lowest order effectively chooses the steady state DF .

The coarse graining series is taken as 
\be
P=P_o+P_1~\alpha^{-1}+P_2~\alpha^{-2}+\dots
\ee
while the fine graining series is assumed in the form
\be
P=P_o+P_{-1}~\alpha+P_{-2}~\alpha^2+\dots
\ee

We apply the coarse-graining expansion in the next sub-section and follow it in a subsequent sub-section with a discussion of the results. A final sub-section in this part applies the fine-graining expansion to show that the `sheet' phase-space structure may be recovered in this way. 

\subsection{Coarse Graining}

We substitute the coarse graining expansion into equation (\ref{isospherVlasov})and obtain at zeroth order the equation
\be
-t\partial_t~P_o+(3\da-1)P_o+(\da)R\partial_R~P_o+(\da-1)Y\partial_Y~P_o=0,\label{coarsezero}
\ee

while the second equation at zeroth order is simply, from equation (\ref{spherPoisson}),
\be
\frac{1}{R^2}\frac{d}{dR}\left(R^2\frac{d~\Psi_o}{d~R}\right)=4\pi G~\Theta_o=4\pi G~\int~P_o~d^3~Y.\label{scaledPoisson}
\ee

Using the method of characteristics in these latter two equations we obtain the solution
\bea
P_o=P_{oo}(\zeta )e^{-(3\da-1)s},\Theta_o=& I_{oo}R^{-(2\ad)},& t=t_o(\zeta)e^{-s},\nonumber\\ 
\Psi_o=-\gamma R^{2(1-\ad)},\zeta &\equiv R^{(\ad-1)}Y,& s=\ad \ln R,
\label{zerothorder}
\eea
where 
\be
\gamma\equiv (4\pi GI_{oo})/(2(3-2\ad)(\ad-1))\label{gamma},
\ee
 and 
\be
I_{oo}\equiv \int~P_{oo}(\zeta)~d^3~\zeta).
\ee

In the preceding formulae, we have assumed that all particles cross $R=1$ at $s=0$ (although not necessarily at the same time of course) with the scaled velocity $Y$ equal to $\zeta$. In general we may identify $s$ with time through $\alpha T=\ln(\alpha t_o)-s$. However there is no assumption of self-similarity until the explicit dependence on $T$ or $t$ is suppressed. We continue for the moment by assuming self-similarity so that $\partial_t=0$, but we will recall below that it is an explicit means of breaking the self-similarity.

In the special case having $\ad=1$ the above formulae apply except that 
\be
\Psi_o=4\pi G I_{oo}\ln(R).\label{isopot}
\ee
It should also be noted that in that case $Y=v=\zeta$.

The first order coarse-grained Vlasov equation now becomes
\be
(3\da-1) P_1+\da R\di R P_1+(\da-1)Y\di Y P_1-Y_R\di R P_o +(\frac{Y_R}{Y})\di R\Psi_o\di Y P_o=0,
\ee
from which we see that the characteristics do not change with coarse grained order and in addition, after a short calculation which takes care to differentiate before evaluating on the characteristics, that

\be
\frac{d~P_1}{d~s}+(3\da-1)P_1=e^{-(3\da s)}(\ad-1)\zeta_R\left((\zeta-2\gamma/\zeta)\frac{d~P_{oo}}{d~\zeta}-\frac{3-\ad}{\ad-1}P_{oo}\right).\label{P_1}
\ee

It is readily seen that all higher order terms in the expansion will vanish if the first order correction vanishes (apart from the homogeneous part which has the same form as $P_{oo}$ and can be absorbed at this order). Thus we obtain the condition 
\be
(\zeta-2\gamma/\zeta)\frac{d~P_{oo}}{d~\zeta}-\frac{3-\ad}{\ad-1}P_{oo}=0,\label{polyDFcoarse}
\ee
to obtain an exact smoothed DF independent of the degree of coarse-graining.

When $\ad=1$ the first order equation becomes
\be
\frac{dP_1}{ds}+2P_1=-\frac{v_R}{R^3}\left(2P_{oo}+\frac{4\pi G I_{oo}}{v}\frac{dP_{oo}}{dv}\right)\label{isoP_1}\label{isoDFcoarse}
\ee

and so the perfectly coarse-grained condition takes the form 
\be
\frac{d~P_{oo}}{d~v}+\frac{2v}{4\pi G I_{oo}}P_{oo}=0.
\ee
Equations (\ref{polyDFcoarse}) and (\ref{isoDFcoarse}) integrate to give respectively
\be
P_{oo}=C|\epsilon_{oo}|^{p(\ad)},
\label{polycoarse}
\ee
and 
\be
P_{oo}=C~e^{-(v^2/4\pi G I_{oo})}.
\label{gausscoarse}
\ee
Here we have set 
\be
p(\ad)\equiv \frac{3-\ad}{2(\ad-1)},
\ee \label{indexp}
and 
\be
\epsilon_{oo}\equiv \zeta^2/2-\gamma.
\ee
This latter quantity is related to the energy $E=\Phi+v^2/2$ by 
\be
E=e^{2(1-\ad)\delta T}~R^{2(1-\ad)}\epsilon_{oo}.
\label{energy}
\ee
Equation (\ref{polycoarse}),together with equations (\ref{zerothorder}) and (\ref{transsphers}), yields an exact, steady (in agreement with Jeans' theorem), coarse-grained DF in the polytropic form  ($E<0$ for convergence)
\be
f=C|E|^{p(\ad)}.
\label{polyDF}
\ee

A similar procedure in the case $\ad=1$ yields the Gaussian
\be
f=C~e^{-\frac{E}{2\pi G I_{oo}}},
\label{gaussDF}
\ee
where$E=v^2/2+4\pi G I_{oo}\ln(r)$.

We conclude that in this phase space geometry, self-similar infall combined with the associated collective relaxation leads to a smoothed DF in the above form. It is generally a polytrope of order 
\be
n=\ad/(\ad-1),\label{index}
\ee
 since the Gaussian form is a limiting polytrope of infinite order which is attained as $\ad\rightarrow 1+$. We expect only the cases $\ad\ge 1$ since otherwise the polytropes are linearly unstable according to the Doremus-Feix-Baumann theorem (DFBT; e.g. Binney and Tremaine, 1987).
 
\subsection {Discussion of Coarse Grained Results}

Although our unique results (\ref{polyDF}) and (\ref{gaussDF}) are found here  for a spherically symmetric isotropic system, we emphasize that similar results may be found using this method for other simple systems such as the 1D motion of collisionless planes, purely radial orbits in spherical symmetry and anisotropic spherical symmetry. They are similar in the sense that a similar sequence of gaussian polytropes is found, but they differ in index depending on the assumed phase space dimensions. This dependence is in accord with the numerical conclusions of Moutarde et al. (1995), Teyssier et al. (1997) and Torman et al. (1997). In addition we show in the appendix that this same sequence is accessible starting from the most general geometry in phase space.

We should note that Merrall and Henriksen (2003) have shown by several independent numerical simulations that even linearly stable polytropes are in fact metastable, and that they collapse after sufficient cooling towards a (lowered) Gaussian distribution, whose temperature increases in time over the initial free-fall value. Given this result, and given that theory as elaborated e.g. by Lynden-Bell (1967), Aly (1989), and Nakamura (2000) suggests that the ultimate DF of a relaxed collisionless system should be a spherically symmetric, isotropic Gaussian, at least sufficiently far from all boundaries; we suggest that the collective relaxation proceeds through a polytropic sequence of distribution functions as above. Moreover since Merrall and Henriksen (ibid) also find that the relaxation `stalls' if there is too little initial symmetry, we suggest further that a system may spend considerable time in a metastable polytropic state before being re-excited by some external encounter and thus being provoked to continue its evolution towards a Gaussian (energy dissipation) or towards a lower order polytrope if the energy exchange is positive. 

We should also recall in this connection that `initially steep' (that is $\ad>1$ in the self-similar regime) halos were shown by Henriksen and Widrow (1997) to be isolated from their surroundings by an external saddle point in the family of possible orbits, and this was suggested as the origin of the `memory' of initial conditions. There is no such point for $\ad <1$ and these evolve rapidly towards $\ad=1$. 
However although suggestive, these results are strictly only for radial orbits.

 In the present case the characteristics of equation (\ref{isospherVlasov}) do not in general admit the same analytic treatment. But in the limiting case of $\ad=2$ this treatment is in fact possible, and one finds, following the procedure of Henriksen and Widrow (1997), that (setting $\alpha=1$ for convenience in this argument) the equivalent one dimensional problem in $R$ becomes
\be
 \frac{d^2R}{ds^2}=-\frac{d\Psi_{eff}}{dR},
\ee
where $\Psi_{eff}\equiv \Psi-R^2/8-J^2/(2R^2)$ and $Y_\perp=J/R$ with $J$ constant. Thus for negative energy particles there is always an outer turning point just as  for purely radial orbits, and in addition there is the inner turning point required by  non-zero angular momentum. Thus  the isolation by critical points is also present in this limit, which is of course initially steep.
Although we can not establish this behaviour for all steep indices this analytic argument adds weight to the numerical evidence for meta-stability cited above. Such behaviour  suggests that the polytropic index will be preserved during the single halo infall and that subsequent disturbances are required to break the meta-stability.

Since no finite system can realize either a global polytrope of $n>5$ or a Gaussian sphere, we expect that the outer parts will be dominated by the edge effect power law $r^{-3}$ (perhaps $r^{-4}$ for tidally truncated cases). In the central bounded regions higher order polytropes are more likely.

But it is the density profile which we must address next. This follows from equations (\ref{zerothorder}) and (\ref{transsphers}) unequivocally as 
\be
\rho=I_{oo}~ r^{-2\ad}\label{density}.
\ee
Moreover the similarity class may be expressed in terms of the primordial density perturbation from which the halo is born (Fillmore and Goldreich, 1985; Bertschinger, 1985; Henriksen and Widrow, 1999) if this is assumed to have been a power law of the form $\delta\rho\propto r^{-\epsilon}$. One finds 
\be
\ad=3\epsilon/2(\epsilon+1),\label{primordindex}
\ee
and so the initial polytropic DF might be expected to have by equation (\ref{index}) the order 
\be
n=\frac{3\epsilon}{\epsilon-2},
\ee
for $\epsilon>2$. 

Thus formally for $\epsilon\ge 2$ we deduce a power law density profile in the bulk of the halo that is approximately $\rho\propto r^{-2}$. This implies a singular density expression of both the polytropic and the gaussian distribution functions. However it can only apply where strict self-similarity is reasonable, which is always in an intermediate region far from boundaries. The polytropic and gaussian distribution functions are compatible with non-singular density distributions as well as with these singular, mathematically isolated, forms and so we must look for evidence that the system will tend to evolve toward the non-singular family. Henriksen and Widrow (1999) have shown that near the edge of such a system one can expect the `Keplerian' power law $\rho\propto r^{-3}$. They were able to give a self-consistent analysis for the region outside a fixed mass, but the simple version of their argument consists of taking $\epsilon$ large as might be expected near the edge of the system, which gives $\ad=3/2$ and hence the Keplerian power law.

So we already expect the inverse square law, to turn over at the edge to become an inverse cube law. This accords qualitatively with the simulations (e.g. Navarro, Frenk and White, 1996) except at the centre of the system.
But it is certainly at the centre of the system where we must break the coarse-graining self-similarity since the expansion will not converge where the potential gradient becomes infinite. We can only hope to approach this region asymptotically by breaking the self-similarity in an effort to find a tendency to flatten. This we do in a subsequent section although the essential idea was already discussed in paper I.

\begin{figure}
\begin{tabular}{cc} 
\rotatebox{0}{\scalebox{0.50} 
        {\includegraphics
                {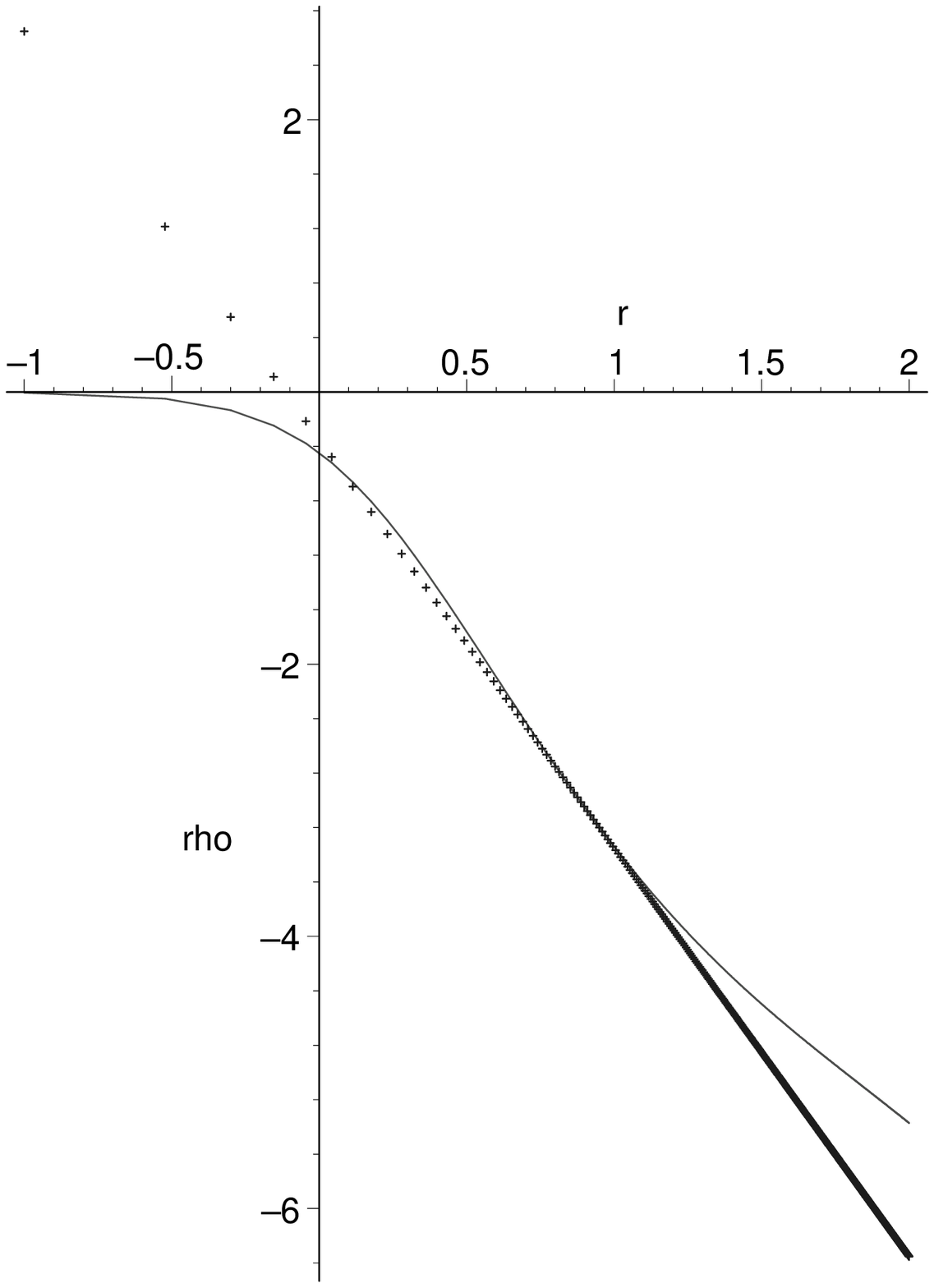}}} &
\rotatebox{0}{\scalebox{0.50}
        {\includegraphics
                {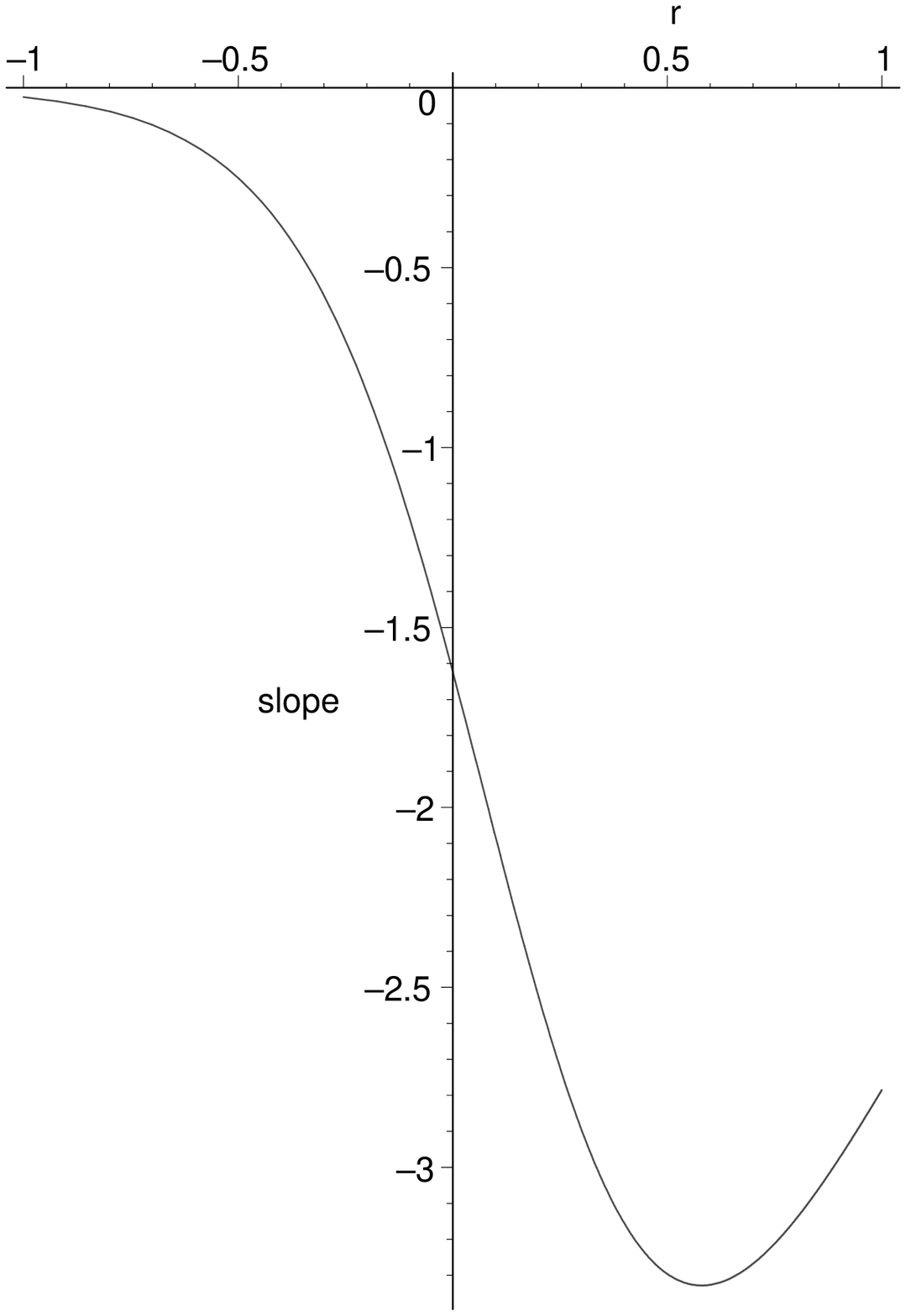}}} \\  
\end{tabular}
\caption{\label{fig:polyfit8}
The left panel shows the log-log density profile of an $n=8$ polytrope together with 
an $r^{-3}$ power law (dotted). The constants in the power law have been chosen so that it fits continuously to the polytrope at $r=7.706$. This is just beyond the point where the mass equals $1$ and is the point where the polytropic slope is $3$. The right panel shows the slope of the polytrope as a function of $\log{r}$.}
 
\end{figure}

Our tentative conclusion is that we should select from the family of polytropes, at least in the central regions to find a continuous prediction for the density profile. We believe it to be a high order polytrope that is fit smoothly to an outer edge effect $r^{-3}$ law. For example as is shown in figure (\ref{fig:polyfit8})  a polytrope of order 8, for which $\ad=8/7$ will connect smoothly a region of slope equal to $-3$ to a central core. As might be expected, even the non-singular isothermal sphere(see e.g. Binney and Tremaine,ibid, p230 ) will join a region of slope nearly equal to $-3$ to a central core. The polytropic region must terminate in any case to keep the mass finite, when $n>5$, and this edge effect is reponsible for the $r^{-3}$ law. Strictly the outer slope should be slightly steeper than this in order to render the mass finite. Our suggested fit to the density profiles of dark matter halos replaces the phenomenological parameters of the NFW fit by the index of the central polytrope. Since this in turn may depend on the age of the system, we expect universal fits only for the oldest systems.
 
One should note however that there are other physical effects that may fix the similarity class in addition to the mass distribution, such as a fixed specific angular momentum, for which $\ad$ attains the limiting value of 2. This is because together with Newton's constant $G$ such a quantity determines a density profile 
\be
\rho=\mu \frac{J^2}{G}~\frac{1}{r^4},\label{Jdens}
\ee
where $\mu$ is a positive numerical constant. We see from equation (\ref{primordindex}) that since we expect $\epsilon$ to be $>2$, there is no value that gives $\ad=2$. Thus we must abandon setting the initial density profile independently of a limiting specific angular momentum $J$. A halo dominated by a constant specific angular momentum  could be relevant to halos formed from material that has been spun up tidally (presumably that which has fallen in from larger spatial scales), and indeed tidally truncated. The outer regions of globular clusters and of elliptical galaxies may reflect this behaviour in their density profiles, and in fact any dark matter halo must ultimately become steeper than $r^{-3}$, probably due to tidal effects. Our argument requires the region of tidal truncation to have a density profile $\rho\propto r^{-4}$.  

Because our transformation to stationary coordinates is canonical in this case, no coarse graining is actually achieved as $\alpha \rightarrow \infty$. However this should not affect the result that is stopped at zeroth order which yields the steady polytrope, since as discussed in the appendix the expansion in $1/\alpha$ can also be understood as a progressive time averaging.

Finally in this section we note that we can determine the positive constant $C$ in equation (\ref{gaussDF}) by introducing  spatial and density scales, and thus completely determine that result in physical terms. We introduce a fiducial density and a fiducial radius $\rho_o$, $r_o$, so that from the Poisson equation we have $ 4\pi G\rho=k/r^2$ and hence  
\be
k\equiv 4\pi G \rho_o r_o^2.
\ee
Then from the integral of the DF in (\ref{gaussDF}) over velocity space (making the usual extention to infinite velocities) we have 
\be
C=(\frac{2}{\pi})^{3/2}\times \frac{1}{4\pi Gr_o^2}\times \frac{1}{\sqrt{k}}.
\ee

In the next sub-section we turn attention to the fine graining limit in order to demonstrate that we can gradually recover the known phase space structure (e.g. Henriksen and Widrow, 1999; Merrall and Henriksen, 2003).

\subsection{Fine Graining}

In this section we proceed by using the same transformation as in the previous section, but now the expansion is of the form
\be
P=P_o+\alpha P_{-1}+\alpha^2 P_{-2}+\dots,
\ee
where $P_o$ is the steady coarse-grained function found in the previous section, see equations (\ref{polyDF},\ref{gaussDF}), and the quantity $\Psi_o$ is the corresponding potential. Recalling the self-similar form (no explicit time dependence) of equation (\ref{coarsezero}), this expansion yields at zeroth order in $\alpha$
\be
Y_R\partial_RP_{-1}-(\frac{d~\Psi_o}{d~R})(\frac{Y_R}{Y})\partial_YP_{-1}=(\frac{d~\Psi_{-1}}{d~R})(\frac{Y_R}{Y})\partial_YP_o.\label{fine-1}
\ee
 From the characteristics of this equation we have 
\bea
\frac{d~R}{d~s}&=&Y_R,~~~\frac{d~Y}{d~s}=-(\frac{Y_R}{Y})(\frac{d~\Psi_o}{d~R}),\nonumber\\
\frac{d~P_{-1}}{d~s}&=&(\frac{d~\Psi_{-1}}{d~R})(\frac{\zeta_R}{\zeta})\partial_YP_o,~~~\frac{Y^2}{2}+\Psi_o=\epsilon.\label{finechar}
\eea
The scaled energy $\epsilon$ is constant on the fine-grained characteristic and it takes the general form 
\be
\epsilon=\epsilon_{oo}~R^{2(1-\ad)},\label{energyfine}
\ee
with $\Psi_o$ either as in equations (\ref{zerothorder}, \ref{gamma}) for $\ad>1$ or as in equation (\ref{isopot}) for $\ad=1$. The coarse grained zeroth order must also be adjusted accordingly when $\ad=1$, becoming $P_{oo}$ from equation (\ref{gausscoarse}) times $R^{-2}$.  Otherwise $\ad=1$ may be substituted directly into the equations above. The zeroth order coarse-grained characteristic $\zeta$ is also as in equation (\ref{zerothorder}).

Proceeding now with the general case of equation(\ref{polyDF}), we calculate from the appropriate member of equations (\ref{finechar}) that
\be
\frac{d~P_{-1}}{d~R}=p(\ad)~R^{3(\ad-1)}~(\frac{P_{oo}(\zeta)}{\epsilon_{oo}(\zeta)})~\frac{d~\Psi_{-1}}{d~R}.\label{P-1}
\ee
We must now use Poisson's equation in the form
\be
4\pi G\Theta_{-1}=\frac{1}{R^2}~\frac{d}{d~R}\left(R^2~\frac{d~\Psi_{-1}}{d~R}\right),\label{Poiss-1}
\ee
with
\be
\Theta_{-1}=\int~P_{-1}~d^3~Y=R^{3(1-\ad)}~\int~P_{-1}~d^3~\zeta,\label{Theta1}
\ee
to write an equation for $\Psi_{-1}$; after which we find $\Theta_{-1}$ and hence $P_{-1}$.

In fact it is easier to solve for $g_{-1}\equiv -d\Psi_{-1}/dR$. We do this by multiplying equation (\ref{Poiss-1}) by $R^{3(\ad-1)}$, substituting equation (\ref{Theta1}) and differentiating with respect to $R$ in order to obtain an equation for $g_{-1}$ in the form
\be
R^2~g''_{-1}(R)+(3\ad-1)R~g'_{-1}(R)+g_{-1}(R)\left((3-\ad)(3-2\ad)\frac{I_{2p(\ad)-1}}{I_{2p(\ad)+1}}~R^2-2(4-3\ad)\right)=0,\label{finebessel}
\ee
where the prime indicates differentiation with respect to $R$. We have introduced the notation ($\Gamma(x)$ denotes the usual factorial or `gamma' function)
\be
I_q\equiv \int_0^{\pi/2}(\cos{\theta})^2~(\sin{\theta})^q~d\theta\equiv (\sqrt{\pi}/4)\Gamma((q+1)/2)/\Gamma((q+4)/2),
\ee
$p(\ad)$ is as defined previously, the polytropic index is $n=p+3/2$, and we note that
\bea
I_{oo}&=&16\pi\sqrt(2)C\gamma^{(p(\ad)+3/2)}~I_{2p(\ad)+1},\label{Ioopoly}\\
A_{oo}&=&16\pi\sqrt(2)C\gamma^{(p(\ad)+1/2)}~I_{2p(\ad)-1}.\label{constants}
\eea
Here $A_{oo}$ is the integral that appears  in the integral over $d^3\zeta$ of $dP_{-1}/dR$, namely
\be
A_{oo}=8\pi\sqrt{2}C~\int_0^\gamma~|\epsilon_{oo}|^{(p-1)}\sqrt{\gamma-|\epsilon_{oo}|}~d|\epsilon_{oo}|.
\ee
We have supposed that $\epsilon_{oo}<0$ for bound particles.

If we recall the definition (\ref{gamma}) for the quantity $\gamma$ then equation (\ref{Ioopoly}) gives the explicit relation between $I_{oo}$ and $C$. 

The solution to equation (\ref{finebessel}) is ($b^2> 0$)
\be
g_{-1}=R^{(1-(3/2)\ad)}~\left(C_1~J_{(3/2)(2-\ad)}(bR)+C_2~Y_{(3/2)(2-\ad)}(bR)\right),\label{sol1}
\ee
where 
\be
b^2=(3-\ad)(3-2\ad)(\frac{I_{2p-1}}{I_{2p+1}})=2\ad.
\ee
From this result and equation (\ref{Poiss-1}) we may find $\Theta_{-1}$ as
\be
4\pi G\Theta_{-1}=-(\frac{2}{R}~g_{-1}(R)+\frac{dg_{-1}(R)}{dR}),
\ee

which should be compared to the zeroth order density from equation (\ref{zerothorder}). 

We see by looking at the asymptotic forms of the Bessel functions near $R=0$, that the second term in equation (\ref{sol1}) varies as $R^{-2}$ there, so that it allows for a point mass. We omit this by proceeding only with the first term, which in turn varies as $R^{(4-3\ad)}$ near the origin. 

\begin{figure}
\epsfig{file=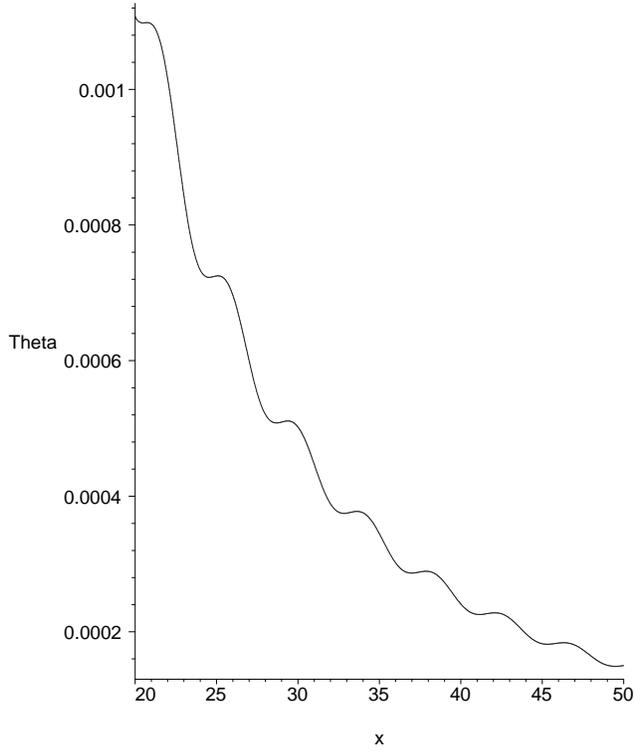,width=0.50\linewidth}
\caption{We show an illustrative plot of the fine structure implied by equation (\ref{sol1}) in terms of $\Theta =\Theta_o+\alpha \Theta_{-1}$ as a function of $x$. $\Theta$ is measured in terms of $I_{oo}/r_o^{2\ad}$, $x\leftarrow r/r_o$, and $\alpha C_1r_o^{0.5\ad}/(4\pi G I_{oo})=0.5$ in order to show clearly the structure. The fiducial radius $r_o$ is arbitrary.}    
\label{fig:finestructure}
\end{figure}

Figure (\ref{fig:finestructure}) shows the variation of $\Theta=\Theta_o+\alpha\Theta_{-1}$ for $\ad=1.125$.  We see that  oscillatory fine structure is detected, which corresponds to the phase sheets seen in the early stages of the  simulations (e.g. Merrall and Henriksen, 2003) before the  the numerical smoothing instability. This shows that the analytic approach  retains the phase sheet structure as indeed it must in the absence of any external perturbations. Of course to completely define the sheets would require higher order terms in the expansion.At this order the fine structure reflects the collective oscillation mode of relaxation. In the non-linear limit we expect these to become the distinct phase sheets. By contrast the coarse-grained result yields  the full chaotic development due to the inevitable presence of numerical uncertainty. 

The limiting case $\ad=1$ must be treated slightly differently. Equations (\ref{finechar}) and (\ref{energyfine}) still apply with the appropriate $\Psi_o$ and $P_o$, but we must remember that $\zeta=Y=v$.  We proceed by calculating 
\be
4 \pi G \Theta_1 =4\pi G ~\int~P_{-1}~d^3 v=-\frac{1}{R^2}\frac{d}{dR}(R^2g_{-1}),
\ee
then taking the derivative with respect to $R$ and using equation (\ref{finechar}) for $dP_{-1}/dR$ to obtain the equation for $g_{-1}$ in the form
\be
R^2\frac{d}{dR}\left(\frac{1}{R^2}\frac{d}{dR}(R^2g_{-1})\right)+\frac{4\pi GA_{oo}}{R^2}g_{-1}=0,\label{isofineosc}
\ee
where now 
\be
A_{oo}=-\frac{1}{2\pi GI_{oo}}\int~v^2\exp{-(v^2/4\pi G I_{oo})}4\pi C~dv,
\ee
and 
\be
I_{oo}=4\pi C \int~v^2\exp{-(v^2/4\pi G I_{oo})}~dv,
\ee
so that
\be
A_{oo}=-1/(2\pi G),
\ee
and 
\be
I_{oo}=\frac{1}{(4\pi^2 G)^3}\frac{1}{C^2}.\label{Ioo}
\ee

Equation (\ref{isofineosc}) is now an Euler equation and the solution takes the form
\be
g_{-1}\propto R^\ell,
\ee
where 
\be
\ell=-1/2\pm i\sqrt{7/4}.
\ee
The general solution thus takes the oscillatory form 
\be
g_{-1}=B~R^{-1/2} \sin{(\sqrt{7/4}\ln{R}+\phi)},
\ee
where $B$ and $\phi$ are arbitrary. 

In this case the oscillations are less pronounced until the divergent centre is approached.  This case is notoriously difficult to calculate in simulations so that it is interesting to establish the continuity with the polytropic behaviour found here.

\section{Breaking the Self-Similarity as $R\rightarrow 0$}

\subsection{Flattening the Singular Isotherm }
 
We break the self-similarity in our method by permitting an explicit dependence on $t$ (Carter and Henriksen, 1991). We begin by showing the method in the algebraically simple case with $\ad=1$. Returning to the section on coarse-graining we can write the zeroth order equation with time dependence as 
\be
t\partial_t~P_o-R\partial_R~P_o-2P_o=0,
\ee
which has the general solution
\be
P_o=\frac{P_{oo}(v,u)}{R^2},
\ee
where $u\equiv Rt$, and the same characteristics as for the first order term below. 

Proceeding to the first order term we find the equation
\be
-t\partial_tP_1+R\partial_RP_1=-2P_1+v_R\partial_RP_o-\frac{d\Psi_o}{dR}\frac{v_R}{v}\partial_vP_o,
\ee
which has the characteristics 
\be
\frac{dt}{ds}=-t;~~~~~~~~~~\frac{dR}{ds}=R,
\ee
and hence 
\be
\frac{dP_1}{ds}+2P_1=\frac{v_R}{R^3}\left(-2P_{oo}-\frac{y}{v}\partial_vP_{oo}+R\partial_RP_{oo}\right).\label{simbreak1}
\ee
Here we define 
\be
y\equiv R\frac{d\Psi_o}{dR},
\ee
for convenience. This quantity is $4\pi G I_{oo}$ in the self-similar coarse-grained solution.
Note that $v$ and $u$ (by direct calculation using the characteristics) are constant on the characteristics.

If we measure $t$ from the onset of the similarity breaking, then by continuity at $t=0$ we expect 
\be
P_{oo}=C(u)\exp{-(v^2/y)}
\ee
since $y\rightarrow 4\pi GI_{oo}\equiv y(0)$ in this limit. We also require $C(0)=C$, the constant appearing in $P_{oo}$ in the self-similar coarse-graining. The equation (\ref{simbreak1}) becomes simply 
\be
\frac{d}{ds}\left(P_1e^{2s}\right)=v_Re^{-s}R\partial_RP_{oo}=v_Re^{-s}P_{oo}(\frac{d\ln{C}}{d\ln{u}}-v^2R\partial_R(1/y)).\label{solbreak1}
\ee
 It is now necessary to write an equation for $ y(R,t)$ using Poisson's equation. 
 From equation(\ref{scaledPoisson}) we have easily 
\be
\frac{d}{dR}(Ry)=4\pi^{5/2}GC(u)(Ry)^{3/2}R^{-3/2},
\ee 
which yields $y=const=y(0)$, $C=const=C(0)$ as one solution as is required at $t=0$. In addition it has the general solution
\be
\frac{2}{\sqrt{Ry}}=-4\pi^{5/2}G~\int^R~\frac{C(R't)}{R'^{3/2}}~dR'-K,
\ee
where $K$ is a constant of integration. 

Since we are interested in the origin of the flattening we proceed by holding $C(u)=C$. It is easy to see from equation (\ref{solbreak1}) that a function $C(u)$ that increases with $u$ will assist any initial flattening, which is presumably due to the wave-particle relaxation itself. This is because a reduced central phase density (decreasing $C$ with decreasing $R$) corresponds to a reduced space density for the same volume of velocity space. 
We are now able to write the solution for $y$  as 
\be
\frac{1}{y}=\frac{1}{4}\left(8\pi^{5/2}GC-KR^{1/2}\right)^2.\label{y}
\ee
We may note here that at a fixed $r$ (with $K>0$ as we require below) $1/y$ increases with increasing time  just as was found in the numerical simulations of Merrall and Henriksen (2003). This time-dependent dispersion is presumably due to the collective relaxation itself. Moreover at large $r$ at any $t$, $1/y$ increases without limit which gives a cosmological DF peaked around zero velocity at large distances. Moreover in this same limit $y\propto 1/R$, which implies a Keplerian potential outside most of the mass at large $r$ just as was assumed and shown to give the $r^{-3}$ exterior density profile in Henriksen and Widrow (1999). 

We now obtain from equations(\ref{solbreak1}) and (\ref{y}) the interesting result
\be
\frac{d}{ds}(P_1e^{2s})=e^{-s}v_Rv^2P_{oo}\left(\frac{4\pi^{5/2}GC}{y^{1/2}}-\frac{1}{y}\right).\label{explicitP1}
\ee
We see that there will be a flattening if the first term in brackets dominates. This must be transitory in time since initially the two terms are equal (remembering equation (\ref{Ioo}), and finally $y$ becomes large compared to the initial value.
If $1/y$ is to be small then by equation (\ref{y}) we must have
\be
K\approx \frac{8\pi^{5/2}GC}{R_f^{1/2}},
\ee
where $R_f$ is the radius where the flattening begins. We now approximate $1/y$ by the first term in equation(\ref{y}) for $R<R_f$ and integrate equation (\ref{explicitP1}) using the dominant first term to obtain
\be
P_1\approx -(4\pi GC)^2\pi^3\frac{v_Rv^2P_{oo}}{R^3}.
\ee
Integrating this result in turn over velocity space (using cylindrical coordinates $d^3v=2\pi v_\perp dv_\perp dv_R$) we find
\be
\Theta_1=-\frac{C}{(GC)^4}\frac{1}{(16\pi^3)^2\pi^3}R^{-3}.
\ee
Consequently to this order
\be
\Theta=\frac{I_{oo}}{R^2}\left(1-\frac{1}{2(2\pi)^3}\frac{1}{\alpha R GC}\right),\label{Thetabreak}
\ee
on using equation (\ref{Ioo}). Now recalling that the physical density is $e^{-2\alpha T}\Theta$ plus the definition of $R$ we obtain the dimensionally correct result
\be
\rho=\frac{I_{oo}}{r^2}\left(1-\sqrt{GI_{oo}}\frac{t}{4r}\right).
\label{flatdensity}
\ee
In this formula $I_{oo}=\rho_or_o^2$ in terms of fiducial values in the self-similar region. It is clear that the flattening increases with dynamical age of the system at a given radius.

\begin{figure}
\epsfig{file=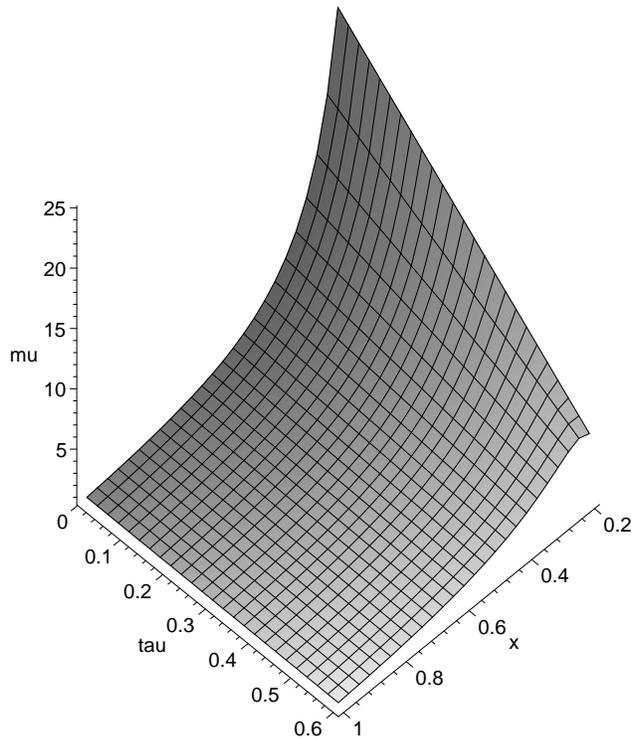,width=0.50\linewidth}
\caption{We show a plot of the scaled density, $\rho/\rho_o$ against $x=r/r_o$ and $\tau=\sqrt{G\rho_o}t$ according to equation (\ref{density}) over the indicated ranges. We see that a core is beginning to form at the `centre' of the system  when $\tau/(4x)=0.75$}    
\label{fig:flat}
\end{figure}

We show in figure (\ref{fig:flat}) an example of the type of flattening this equation implies. The figure shows $\rho/\rho_o$ against $x=r/r_o$ and $\tau=\sqrt{G\rho_o}~t$. We observe that by the time that $\tau/4x$ reaches $0.75$ a central core is forming.  
 
There is no other physics in the problem but the collective relaxation provided by the wave-particle scattering. The time-scale of the density evolution seen in equation (\ref{flatdensity}) is essentially a few crossing times at $r$.

However we know from numerical work (e.g. Merrall and Henriksen, 2003) that sufficiently asymmetric initial systems will not be fully relaxed before the small scale time dependence is damped away, presumably by Landau damping. Thus the time that appears above is measured from each new epsisode of relaxation, that may be stimulated by mergers. For this reason and because of the demonstrated meta-stability of equilibrium polytropes, we believe that the relaxation can proceed through the polytropic sequence that we have found in (\ref{polyDF}). 

For this reason we discuss briefly in the next sub-section the similarity breaking for this sequence.

We note that low mass systems are older in CDM cosmology and so they may be expected statistically to be more relaxed. This is the basis of a possible statistical test of equation (\ref{flatdensity}), but this is best left to a separate paper.

\subsection{Flattening the Singular Polytropes}

Since the method is precisely that of the previous section, we content ourselves here with giving a catalogue of the analogous formulae. We will discuss the results carefully however as they are somewhat surprising, due principally to the division at $n=5$ ($\ad=5/4$) between spatially finite and spatially infinite systems.

The zeroth order scaled DF is found to be 
\be
P_o=P_{oo}(\zeta,u)R^{-(3-\ad)},\label{polyPo}
\ee
where 
\be
u\equiv Rt^\da,~~~~~~\zeta\equiv YR^{(\ad-1)},\label{polyvars}
\ee
and the coarse-grained characteristics are 
\be
R=e^{\da s},~~~~~~dt/ds=-t.
\ee

The equivalent of equation (\ref{solbreak1}) for the first order scaled DF is
\be
\frac{dP_1}{ds}+(3\da-1)P_1=\frac{Y_RP_{oo}}{R^{4-\ad}}\left(\frac{d\ln{C(u)}}{d\ln{u}}-\frac{p(\ad)}{\epsilon_{oo}}R\partial_R\gamma\right),\label{solbreak2}
\ee 
where by continuity at $t=0$ as above 
\be
P_{oo}=C(u)|\epsilon_{oo}|^p,
\ee
and 
\be
\gamma=\frac{1}{2(\ad-1)}R^{(2\ad-1)}\frac{d\Psi_o}{dR}.
\ee
Formally here
\be
\epsilon_{oo}=\frac{\zeta^2}{2}-\gamma(R,t),\label{eoo}
\ee
just as in the self-similar coarse-graining, but now we must solve for $\gamma$ as in the previous section. For convergence we work as usual with bound systems for which $\epsilon_{oo}\le 0$.

Using the Poisson equation in zeroth order and proceeding as above we find
that $\gamma$ satisfies 
\be
\frac{dw}{dR}=\frac{32\sqrt{2}\pi^2}{\ad-1}I_{2p+1}\frac{GC(u)}{R^{(2-\ad)/(\ad-1)}}~w^{n(\ad)},\label{polygamma}
\ee
where $w\equiv \gamma(R,t)R^{(3-2\ad)}$ for brevity and $n(\ad)$ is as in equation (\ref{index}). 

When $C(u)=C(0)=C$, this equation has the solution $w=\gamma R^{(3-2\ad)}$ with $\gamma=(2\pi G I_{oo})/((\ad-1)(3-2\ad))$ as required. In general the solution may be written (assuming $C$ constant) in the form 
\be
\frac{1}{\gamma}=\left(\frac{32\sqrt{2}\pi^2I_{2p+1}}{(\ad-1)^2q}GC-KR^{q(\ad)}\right)^{(\ad-1)},\label{solpolygamma}
\ee
where
\be
q\equiv \frac{3-2\ad}{\ad-1}.
\ee

When $\ad>5/4$ the system must have an `edge' and so we set $KR_E^q$ equal to the constant in the bracket of equation(\ref{solpolygamma}), where $R_E$ is the current edge of the system. Note that $K>0$ and that $\gamma\rightarrow constant$ as $R\rightarrow 0$. 

When $\ad<5/4$, the solution extends to infinity where $\gamma$ must go to zero so that we must take $K<0$ in equation (\ref{solpolygamma}). We note that in this case there will  be a radius given by $KR_E^q\approx$ bracket constant, {\it outside of which} $\gamma$ is dominated by the second term in the in the bracket.

We are now able to perform the calculation indicated in equation (\ref{solbreak2}) and so find
\be 
P_1=-\frac{(3-\ad)(3-2\ad)}{2(3-4\ad)}\frac{Y_RP_{oo}}{|\epsilon_{oo}|}K\gamma^nR^{(-7+\ad n)}.\label{polyP1}
\ee
Integrating this over $Y_Rd^3Y=R^{4(1-\ad)}2\pi\zeta_Rd\zeta_R\zeta_\perp d\zeta_\perp$ yields finally the `flattening' first order scaled density in the form 
\be
\Theta_1=-\frac{2^{p+1}(\ad-1)}{q}\pi C K~\gamma^{\frac{3\ad+1}{2(\ad-1)}}~R^{(-3+\ad\frac{4-3\ad}{\ad-1})}.\label{polyflatdensity}
\ee

We  remark that for the finite polytropes  as $R<R_E$  $\gamma$ is constant, and since $K>0$ we have flattening as $R\rightarrow 0$ provided that the power of $R$ in $\Theta_1$ is more negative that $-2\ad$, which is the power of the zeroth order singular density. This requires $\ad>(\sqrt{13}-1)/2\approx 1.303$, which corresponds to $n<4.303$. Between this value and $\ad=5/4$ there is no asymptotic flattening as $R\rightarrow 0$, although there may be a `shoulder' effect at $R$ between $R_E$ and $0$. Recalling that $R=re^{-\delta T}=r/(\alpha t)^{\da}$, we see that such a shoulder region expands in time. 

When $\ad<5/4$ there is a regime $R>R_E$ where $\gamma$ is dominated by the second term in (\ref{solpolygamma}). Then in equation (\ref{polyflatdensity}) $\gamma$ is replaced by $|K|$ and the power of $R$ becomes
\be
\Theta_1\propto R^{-(3+p)},
\ee
and so there is always flattening in this regime, which is very rapid as $\ad\rightarrow 1$ since then $p\rightarrow \infty$. This flattening will cease however as $R\rightarrow R_E$ from above, so that once again it is only detectable in this order in a shoulder region. This may be related to the fact that polytropes have a shoulder domain outside of the core where the local slope is less than $-2\ad$. Thus we detect a flattening relative to the self-similar slope beyond this domain. The same effect is present in the isothermal case discussed above since $1/y$ ultimately becomes `large' at small $R$ and the flattening vanishes ($1/y$ in equation (\ref{explicitP1}) dominates the first term).

We conclude from this analysis that the singular isothermal and polytropic spheres will tend towards the more general family of non-singular isothermal and polytropic spheres. This is supported both by the direct calculation of flattening found in this section as a result of the broken self-similarity, and by numerical simulations (Merrall and Henriksen, 2003). Our analytic calculations conclude correctly that a steep region exterior to the core develops as part of this evolution.

\section{Conclusions}

In this paper we have focussed on the precise results that may be found by the resolution expansion method for spherically symmetric, isotropic collisionless systems. By insisting that the coarse-grained equilibrium structure should have no sub-structure (above the scale where the DF is well-defined) we show that the self-similar infall of collisionless matter assumes either a polytropic DF or a gaussian DF depending on its similarity class. These are compatible with the similarity density profiles that are strongly cusped at the centre of the system. 

However these singular polytropes (including the isotherm) are isolated mathematically in the family of mass distributions that are consistent with the polytropic DF. They are held in place so long as self-similarity is assumed by the pre-assigned `class'. We have therefore studied the explicit breaking of the self-similarity that must occur at the centre of the system. This study shows that in lowest order the density profile tends to flatten relative to the self-similar value, at least in a `shoulder' region outside the true core. The flattening is progressive in time, and we conclude that this will be more evident in dynamically old systems such as dwarf galaxies, rather than young systems such as massive clusters. However the time-scale of the flattening is only several crossing times since it is due to the collective wave-particle relaxation, so that each system should be considered on its merits. Since there is a link between the statistical age of a system and its mass in CDM cosmology, this effect should be testable. We leave this to a less pedagogical paper.

Our major suggestion is that this flattening, together with the evolution towards a central gaussian DF shown in numerical simulations (Merrall and Henriksen, 2003) for destabilized polytropes, should lead to the development of a central polytrope in collisionless matter.  The polytropic index is given in equation (\ref{index}) in terms of the similarity class. We expect this to be larger than $1$ since both in principle and in practice the inverse case is unstable to rapid evolution towards $1$. The index may be specified in terms of the initial perturbation that created the infall, or perhaps in terms of  other dominant effects such as tidal truncation.

This modifies slightly the theoretical expectations of Lynden-Bell and Nakamura (ibid), but this is probably because the polytropic state is only intermediary. Systems may  stall in a polytropic state (or indeed one that contains evident sub-structure) since it seems that sufficiently asymmetric initial configurations cease relaxing collectively before a true gaussian is attained (Merrall and Henriksen, ibid). We believe that subsequent disturbances to the system such as those provided by mergers, can restart the collective relaxation and move the system farther along the road to equilibrium. This must be tested in detail however. 

We have pointed out both here and in Henriksen and Widrow (1999) that there is an inevitable edge effect in these systems when one is outside most of the mass. This `law of the edge' that we also refer to as `Keplerian' is simply $\rho\propto r^{-3}$ when the outer regions are dominated by a primordial density power law. However faster cutoffs are possible if for example tidal truncation plays a role ($\rho\propto r^{-4}$ in the limit), and indeed some such regime is essential if the halo is to be of finite mass. 

The fitting function that we recommend for the density profile of a not too evidently unrelaxed system (i.e. if the system were evidently bipolar one would consider each part separately) is shown in figure (\ref{fig:polyfit8}). The phenomenological parameters of other fitting functions are replaced here by the index of the central polytrope, since the fit to the law of the edge must take place where the slope of the polytrope is $-3$ (see figure). This index is in turn a direct link to the initial conditions of the dark matter infall. The degree of flattening is a direct link to the dynamical age of the system.

The density profile for the metastable polytropes is found  from the Lane-Emden equation, assuming that they flatten in accordance with their distribution functions inside the power law, self-similar range. The high order polytropes ($n>5$) have to be terminated arbitrarily to avoid infinite mass, just as does the non-singular gaussian.

It is interesting that in a recent paper by Roy and Perez (2004), wherein the evolution towards equilibrium of a collapsing system of N non-interacting particles is studied by means of a tree-code, they find that the equilibrium density profiles are fitted equally well by a high order polytrope (the Plummer sphere) or by a non-singular isothermal sphere.

All of this behaviour seems to be a direct result of time dependent relaxation that is inevitable in such a system, just as was envisaged in the term `violent relaxation' by Lynden-Bell (ibid) long ago. Of course the relaxation is effective without really changing the particle energies `violently', and so we prefer to refer to it as `collective relaxation' or `wave-particle interaction'. This latter description evokes the Landau acceleration and damping that must be occurring. In any event we do not need physics beyond that of collisionless self-gravitating matter to understand at least the qualitative nature of the dark matter halo density profiles. The basic mechanism appears to be the retarded infall of the more massive shells for $\ad<3/2$. The appendix shows that this effect is not restricted to spherical symmetry. 

In addition to these practical conclusions we have retained a number of pedagogical aspects in the present paper. We retain the mathematical details in order to be as convincing about the utility of the resolution expansion method as possible. In addition we have included a section on fine-graining, where we show that in lowest order the phase space structure found in numerical studies is partially recovered.  This sub-structure appears as outward propagating waves in linear order ($R=r/(\alpha t^\da)$), which we take as an interesting confirmation of the `wave-particle' nature of the interaction. In any case it shows that the expansion method is rather flexible and reliable. It is in fact more general than in this application to self-similar infall, since any parametric non-canonical transformation may be used (e.g. paper I)

\begin{figure}
\begin{tabular}{cc} 
\rotatebox{0}{\scalebox{0.25} 
        {\includegraphics
                {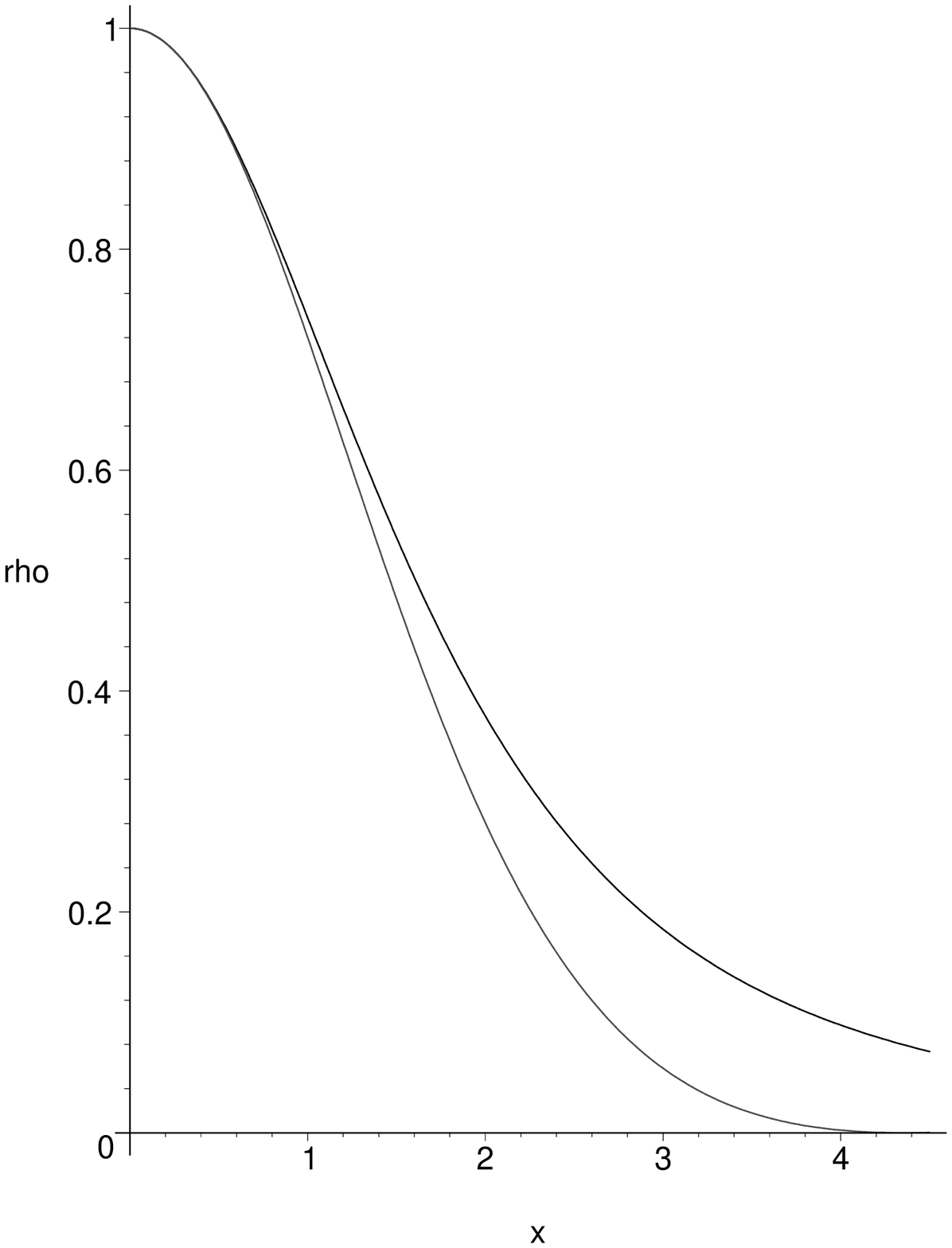}}} &
\rotatebox{0}{\scalebox{0.25}
        {\includegraphics
                {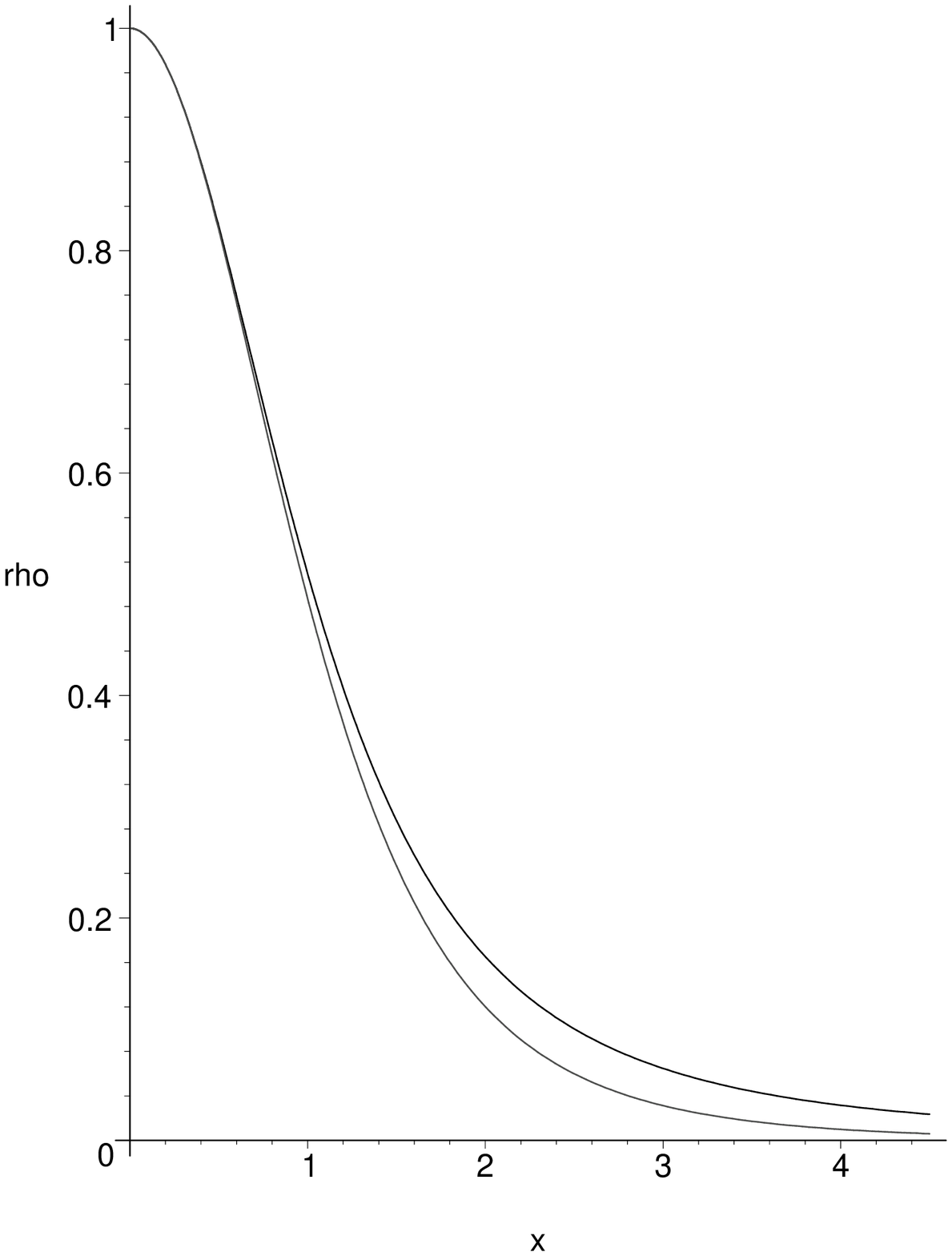}}} \\  
\rotatebox{0}{\scalebox{0.25}
        {\includegraphics
                {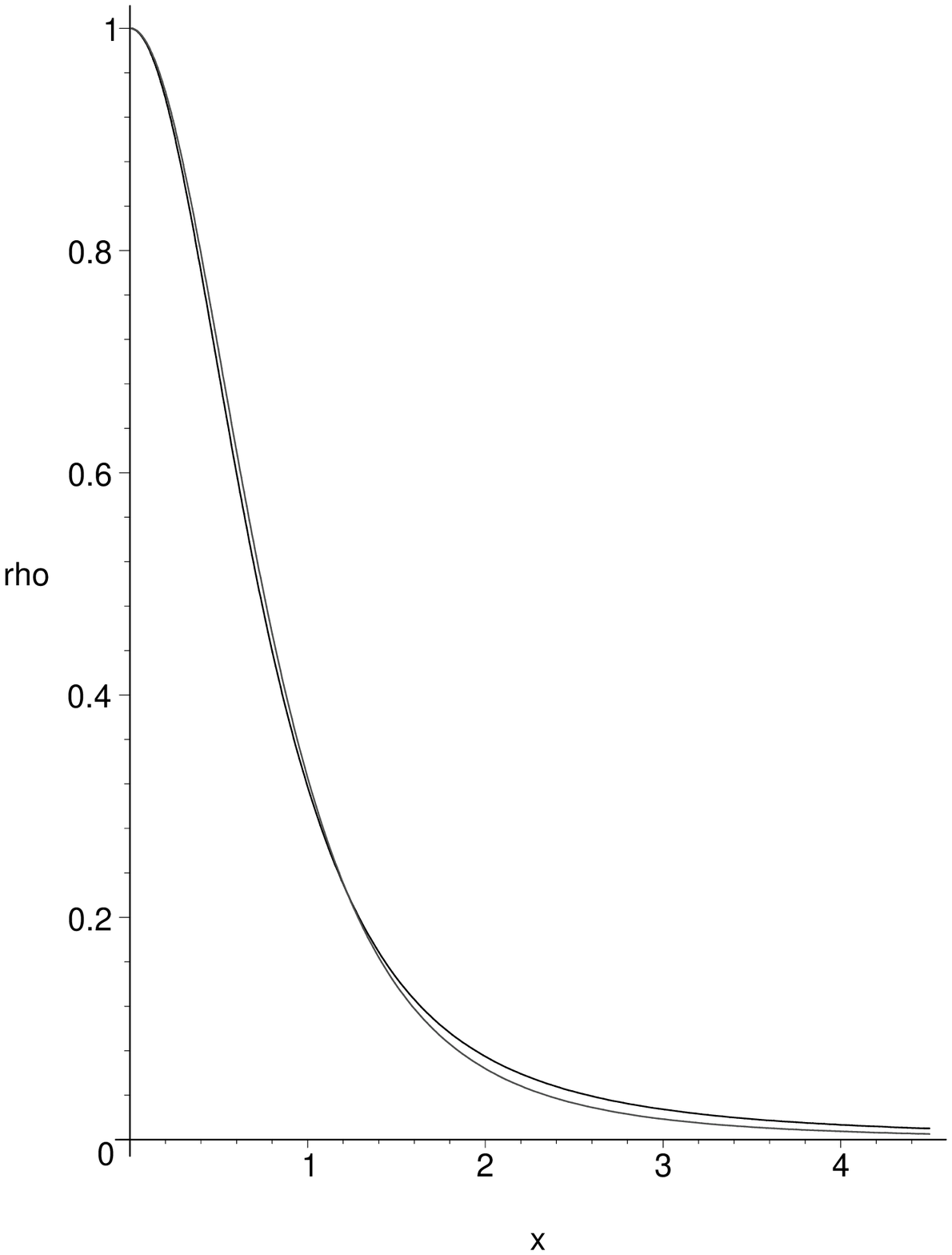}}} &     \\
\end{tabular}
\caption{\label{fig:poly1}
We show on the left panel the density profile of an $n=2$ polytrope, with an isothermal sphere superposed (always the curve with the larger asymptotic density) that has a velocity dispersion equal to $0.5\Psi_o$. The densities are normalized to the central values.
 The right panel shows the same information for an $n=5$ polytrope, with an isothermal sphere superposed that has a velocity dispersion equal to $0.2\Psi_o$.There is little distinction near the centre of the system. The scaling used is given in the text. The lower left panel shows an $n=9$ polytrope with a superposed Gaussian having a dispersion equal to $0.1\Psi_o$} 
\end{figure}

We show in figure (\ref{fig:poly1}) for reference the density profiles for the $n=2$ polytrope (corresponding to a constant angular momentum including zero and $\ad=2$) and for an $n=5$ polytrope (a Plummer sphere). Isothermal spheres that fit well with the polytropes at the centre of the system are superposed. We have measured the densities in terms of a common central density $\rho_o$ and we have used the same scale factor $b^2=1/(4\pi G \Psi_o^{(n-1)}c_n)$, in the notation of Binney and Tremaine (ibid). This requires $\Psi_o=4\pi G\rho_o b^2$ and 
$\Psi_o^n=\rho_o/c_n$. Thus with $\rho_o$ and $b$ chosen the curves are determined physically. The remaining parameter in the equations is $\Psi_o/\sigma^2$, and we show the isothermal curve for $\Psi_o/\sigma^2=2$ together with the $n=2$ polytrope and that with $\Psi_o/\sigma^2=5$ together with the Plummer sphere. The profiles are not easily distinguished near the centre of the system. When $n=9$ the curves can be made to fit over a wide range as shown.

In an appendix we present the ambiguities that arise in a study of a completely general system. There is always an access path to the polytropic/gaussian state 
however, and general arguments due to Lynden-Bell, Nakamura (ibid) and Aly (1989) allow similar conclusions to those found for this special case.

\section{Acknowlegements}

I wish to acknowledge the award of a discovery grant from the Canadian NSERC which partially supported this work, as did the award of sabbatical leave by Queen's University at Kingston. In addition the hospitality and support provided by the Service d'Astrophysique du CEA Saclay (France) allowed the completion of this manuscript. I am grateful to a thorough and perceptive referee for suggestions that improved the manuscript.

\newpage

\appendix{{\bf Appendix: Collisionless Systems in 3D with no imposed symmetries}}

\vskip 1 true cm

In this appendix we contrast the behaviour of an accreting halo that is permitted to fill a full six-dimensional phase space, with the restricted halo studied in the body of the paper. 

The relevant equations are those of section (\ref{sec:geneqs}), before the restriction to the spherically-symmetric isotropic case.

We proceed to the expansion in terms of reciprocal powers of $\alpha$ as usual, but we note from the variable transformations that this represents a coarse graining of phase space only if $\ad<2$. Should $\ad=2$ we have an interesting special case where there is a global integral having the dimensions of angular momentum. Moreover the transformation is canonical in this limit so that there is no change in resolution as $\alpha$ increases. We may still expand in a series in $\alpha^{-1}$ however and this corresponds (see the time transformation in\ref{transsphers}) at a fixed $T$ to a cruder resolution in time ($\Delta t=e^{\alpha T}\Delta T$). We know from Fillmore and Goldreich (1984) and Bertschinger (1985) that if the halo develops from a primordial density perturbation characterized by an effective power law $\propto r^{-\epsilon}$, then the similarity class is given by (\ref{primordindex}) in the text. 

Since we expect $\epsilon$ to be positive there is no value that gives $\ad=2$. Thus we must abandon setting the initial density profile independently of a limiting specific angular momentum $J$. With Newton's constant $G$ this determines a density profile (\ref{Jdens}). This may well be relevant to halos formed from material that has been spun up tidally (presumably that which has fallen in from larger spatial scales). The outer regions of globular clusters and of elliptical galaxies may reflect this behaviour in their density profiles.  

That specific angular momentum is not always conserved  in spherically symmetric potentials (see below) is due to the time dependence of the potential in these non-isolated systems. The mass growth inside a given elliptical orbit is not adiabatic. 
 
Proceeding with the expansion the zeroth order DF may be found by the method of characteristics to be (we drop once again the explicit time dependence, which is the condition for self-similarity) 
\bea
P_o=P_{oo}(\vec\zeta,\theta,\phi,R_o )e^{-(3\da -1)s},~~~R&=& R_o e^{\da s},~~~ \vec\zeta=\vec Y R^{(\ad-1)},\nonumber\\ 
                                            \Theta_o&=& \int~P_o~d^3Y,
\label{charvars}
\eea 
where $\vec\zeta$ and $\theta,~\phi,~R_o$ are constants on the characteristics. We suppose for simplicity that all characteristics cross the same $R=R_o$ so we may set this equal to $1$. This is most likely to be a (scaled) radius that each particle crossed on entering the system since it is clear subsequently that particles are trapped near the centre of the system and we are forbidden to take $R_o$ to be arbitrarily small. It is clear that this choice determines the origin on each characteristic although each particle is in general present at $s=0$ at very different times. In this case we have 
\be
s=\ad\ln{R},
\ee
and the zeroth order DF as 
\be
P_o=P_{oo}(\vec\zeta,\theta,\phi)R^{(\ad-3)}.
\label{3DspherDF}
\ee

The formula for the scaled density becomes
\be
\Theta_o=R^{-2\ad}\int~P_{oo}~d^3\zeta\equiv I_{oo}(\theta,\phi) R^{-2\ad}.
\ee

It is worth emphasizing here that whenever the series is stopped at the zeroth order we have a steady solution; since by definition $\vec\zeta\equiv \vec v r^{(\ad-1)}$, which is time independent along with $\theta$ and $\phi$, and since from the various definitions 
\be
f_o=P_o e^{(\alpha-3\delta)T}=P_{oo}(\vec\zeta,\theta,\phi) r^{-(3-\ad)}.\label{SphoDF}
\ee
 Another way of looking at the result is to note that $\alpha$ is effectively infinite when the series is exact at zeroth order, and thus there can be no time dependence since such an $\alpha$ implies an average over all time.

To continue the expansion to higher orders we must consider the solution of the Poisson equation in zeroth order. This is 
\be
\nabla^2\Psi_o=4\pi G R^{-2\ad}I_{oo}(\theta,\phi).
\ee
One form of the solution in spherical harmonics is 
\be
\Psi_o=\frac{4\pi G}{i(\ad)}R^{(2-2\ad)}A(\theta,\phi),\label{Poiss3D}
\ee
where $i(\ad)\equiv (2-2\ad)(3-2\ad)$. Here the function $A$ is given by 
\bea
A=\Sigma~A_{lm} Y_{lm}(\theta,\phi),& A_{lm}=\frac{i(\ad)~I^*_{lm}}{i(\ad)-\ell(\ell+1)},\nonumber\\
I^*_{lm}&\equiv \int~I_{oo} Y^*_{lm}~d\Omega.
\eea
The spherical harmonics $Y_{lm}$ are used with the convention adopted in Binney and Tremaine (1987) and must not be confused with the scaled velocity used above. The superscript $*$ indicates complex conjugate and $\ell$, $m$ are the usual integers. The value $\ell=2-2\ad$ must be excluded, and we note for future reference that $A/i(\ad)$ is finite and negative ($\ell=0,~\ell=-1$ are excluded) as $i\rightarrow 0$ or $\ad\rightarrow 3/2$. 

The procedure is slightly different in the excluded case of $\ad=1$. A sum separation yields a general result (a product separation requires a proportionality between $A$ and $I_{oo}$) in the form
\be
\Psi_o=k\ln R+A(\theta.\phi),\label{isoPoiss3D}
\ee
where $A(\theta,\phi)$ satisfies 
\be
\frac{1}{\sin{\theta}}\partial_\theta(\sin{\theta}\partial_\theta A)+\frac{1}{\sin^2{\theta}}\partial_\phi^2 A=4\pi G I_{oo}-k. \label{potiso}
\ee
Here $k$ is a true constant.

Proceeding now to the first order in the expansion we obtain for the next term in the DF (taking care to differentiate before evaluating on the characteristics (such that $\vec\zeta$ is $R$ dependent) we find that
\bea
\frac{d}{ds}\left(P_1 e^{(3\da-1)s}\right)e^{-(3\da-1)s}=~~~~~~~~~~~~~~~~~~~~~~&\nonumber\\
R^{-3}\{-(3-\ad)\zeta_R P_{oo}+(\ad-1)\zeta_R\vec\zeta.\partial_{\vec\zeta}P_{oo}+ \zeta_\theta\partial_\theta P_{oo}+\frac{\zeta_\phi}{\sin{\theta}}\partial_\phi P_{oo}+(\zeta_\theta^2+\zeta_\phi^2)\partial_{\zeta_R}P_{oo}&\nonumber\\
-2(1-\ad)(4\pi G A/i(\ad))\partial_{\zeta_R}P_{oo}+(\zeta_\phi^2\cot{\theta}-\zeta_R\zeta_\theta-4\pi G\partial_\theta A/i(\ad))\partial_{\zeta_\theta}P_{oo}&\nonumber\\
-(\zeta_\phi\zeta_R+\zeta_\phi\zeta_\theta\cot(\theta)-\frac{4\pi G}{i(\ad)\sin(\theta)}\partial_\phi A)\partial_{\zeta_\phi}P_{oo}\}&.\label{P1}
\eea                         
This result is of interest in itself since, as has been discussed in paper I, it can be used to describe the deviation from the self-similar density law near $R=0$. However for our present purposes 
setting this expression equal to zero yields a linear partial differential equation for the exact steady $P_{oo}$ in the intermediate self-similar domain as in the examples above, and this may in turn be converted to the following holonomic system of equations using the method of characteristics:
\bea
\frac{dP_{oo}}{da}&=& (3-\ad)\zeta_R P_{oo},\nonumber\\
\frac{d\zeta_R}{da}&=& (\ad-1)\zeta_R^2+\zeta_\theta^2+\zeta_\phi^2-4\pi G A/(3-2\ad),\nonumber\\
\frac{d\zeta_\theta}{da}&=&(\ad-1)\zeta_R\zeta_\theta+\zeta_\phi^2\cot(\theta)-\zeta_R\zeta_\theta-4\pi G\partial_\theta A/i(\ad),\nonumber\\
\frac{d\zeta_\phi}{da}&=& (\ad-1)\zeta_\phi\zeta_R-\zeta_\phi\zeta_\theta\cot(\theta)-\zeta_\phi\zeta_R-4\pi G\partial_\phi A/i(\ad)(\sin{\theta}),\label{gsphch}\\
\frac{d\theta}{da}&=& \zeta_\theta,\nonumber\\
\frac{d\phi}{da}&=& \zeta_\phi/\sin{\theta}.\nonumber
\eea
For the excluded case with $\ad=1$ the same procedure using the zero order potential of equation (\ref{isoPoiss3D}) yields (we write them again for convenience but note that it suffices to replace $4\pi G/i$ by $1$ and $4\pi G A/(3-2\ad)$ by $k$ in addition to $\ad=1$ to obtain the following)
\bea
\frac{dP_{oo}}{da}&=& 2\zeta_R P_{oo},\nonumber\\
\frac{d\zeta_R}{da}&=&\zeta_\theta^2+\zeta_\phi^2 -k,\nonumber\\
\frac{d\zeta_\theta}{da}&=&\zeta_\phi^2\cot(\theta)-\zeta_R\zeta_\theta-\partial_\theta A,\nonumber\\
\frac{d\zeta_\phi}{da}&=&-\zeta_\phi\zeta_R-\zeta_\phi\zeta_\theta\cot(\theta) -\partial_\phi A/(\sin{\theta}), \label{sphchiso}\\
\frac{d\theta}{da}&=& \zeta_\theta,\nonumber\\
\frac{d\phi}{da}&=& \zeta_\phi/\sin{\theta}.\nonumber
\eea
Here $a$ is taken to be the `path parameter' in the phase space of this equation, but it will normally not appear in physical quantities.

These equations may be rearranged in two important ways. In the first instance we can combine the  third and fourth of these equations to obtain 

\be
\frac{1}{2}\frac{d\zeta_\perp^2}{da}=-(2-\ad)\zeta_R\zeta_\perp^2-\frac{4\pi G}{i}\frac{dA}{da}
\label{res1}
\ee
whence on eliminating $\zeta_R$ using the first of these equations one finds 
\be
\frac{d\ln{(P_{oo}(\zeta_\perp^2)^{q(\ad)})}}{da}=-\frac{4\pi G(3-\ad)}{i(\ad)(2-\ad)}~~\frac{1}{\zeta_\perp^2}~\frac{dA}{da},\label{res21}
\ee
where we have used the notations
\be
\zeta_\perp^2\equiv\zeta_\theta^2+\zeta_\phi^2,~~~~q(\ad)\equiv \frac{(3-\ad)}{2(2-\ad)}.\label{notes}
\ee
 One simply sets $4\pi G/i=1$ in these equations in addition to $\ad=1$ in order to obtain the result for the excluded (isothermal) case. We observe that there is never an integral unless $dA/da=0$, which holds strictly only in spherical symmetry . 

However and more importantly, considering for the moment the general case, the last five equations of (\ref{gsphch}) may all be combined to give 
\be
\frac{d\ln{(\frac{\zeta_R^2+\zeta_\perp^2}{2}+\frac{4\pi G A}{i(\ad)})}}{da}=-2(1-\ad)\zeta_R,
\ee
and this in turn may be combined with the first equation to write the the steady, stable, self-similar DF in the form
\be
P_{oo}=B(\kappa_i)|e|^{p(\ad)}.\label{genDF1}
\ee
Here $\kappa_i$ denotes any other integrals, the index $p(\ad)$ is  as in equation (\ref{indexp})) of the text.

and the scaled specific energy $e$ is 
\be
e=\frac{(\zeta_R^2+\zeta_\theta^2+\zeta_\phi^2)}{2}+\frac{4\pi G A(\theta,\phi)}{i(\ad)}.\label{scaen}
\ee

If we now reconstruct the physical DF from the various definitions we find that \be
f=B(\kappa_i)~|E|^{p(\ad)},\label{genDF2}
\ee
where the specific energy is 
\be
E=\frac{(v_r^2+v_\theta^2+v_\phi^2)}{2}+\Phi(r,\theta,\phi),
\ee

if the potential is 
\be
\Phi=\frac{4\pi G A(\theta,\phi)}{i(\ad)}r^{(2-2\ad)}.
\ee
 In general any $\kappa_i$ will be constructed from quantities invariant on the characteristics of the CBE so that Jeans' theorem is satisfied. We have not in fact found any other explicit integrals in the general case as might be expected in the absence of any phase space symmetry. If such integrals exist they are time independent since all of the variables in equation(\ref{gsphch}) are themselves time independent. 

We should note that this DF is the only form that is steady in zeroth order independently of the higher orders and it may in fact be deduced this way (cf HW99). However the present procedure shows explicitly that it is a stable self-similar form that may be compared to the isotropic result. It is more general in terms of the gross asymmetry that is permitted. In this case the form is not obviously unique, assuming that other integrals exist. In the absence of any imposed symmetry however it seems likely to be relevant with $B=b$ a constant. This constant has to be determined self-consistently with the Poisson equation of course.

Proceeding in the same fashion with the last five of equations (\ref{sphchiso}) for the case $\ad=1$ we find directly that 
\be
P_{oo} =B(\kappa_i)~e^{-\frac{2~e}{k}},
\ee

where the scaled specific `energy' is (the radial part of the potential does not yet appear) 
\be
e=\frac{\zeta_R^2+\zeta_\perp^2}{2}+A(\theta,\phi).
\ee
   
This leads through the various definitions to the steady, stable, self-similar DF in this isothermal case as
\be
f=B(\kappa_i)~e^{-\frac{2E}{k}},\label{DFiso}
\ee
where the specific energy is 
\be
E=\frac{v_r^2+v_\theta^2+v_\phi^2}{2}+A(\theta,\phi)+k\ln r.
\ee

Once again we do not find any other explicit integrals, and since this is the gaussian form expected on theoretical grounds (Lynden-Bell 1967, Nakamura, 2000) we suspect it is the unique form. It can also be deduced by requiring a steady DF for $\ad=1$ without the present formalism. It is an important result in any case as it shows that the density profile ($1/r^2$) that appears so frequently in numerical simulations is also associated with the gaussian DF expected on theoretical grounds (ibid) in general symmetry.

It should also be noted that we have omitted discussing the final phase of any specific choice of DF, which is to solve the resulting Poisson equation for the potential function $A(\theta,\phi)$. If we consider only what we suppose to be the limiting case when $\ad=1$, then we have to solve the isothermal type potential equation (\ref{potiso}) with 
\be
I_{oo}=e^{-\frac{2A(\theta,\phi)}{k}}~\int~B(\kappa_i)~e^{-\frac{\zeta^2}{k}}~d^3\zeta.
\ee
 
Even with $B$ constant, we see that there may be only certain values of $k$ that are permitted (for example $k=\lim_{\theta\rightarrow 0}{4\pi G I_{oo}}$ in order to avoid a singularity on the axis). One solution is clearly $A=constant$ and $k=4\pi G I_{oo}$, that is spherical symmetry, which is clearly a minimum energy condition for positive $A$.

The special case $\ad=2$ is also of interest as remarked above.The equations (\ref{gsphch}) now yield 
\be
f=B(\kappa,\kappa_i)|E|^{1/2},\label{canDF}
\ee
in agreement with equation (\ref{genDF2}), but now in addition we have the integral
\be
\kappa=\frac{\zeta_\perp^2}{2}+2\pi G~A(\theta,\phi),\label{canDFI}
\ee 
which certainly removes the strict uniqueness of the DF. The variables $\vec\zeta$ here have the dimensions of `actions', namely $\propto r\vec v$. The radial dependence of the potential  is $r^{-2}$ so that we are strongly forbidden to continue this case toward the centre of the system. Nevertheless this case illustrates nicely the fact that with more `isolating ' integrals the DF is less unique. It seems that it is best defined in the absence of integrals in addition to the energy, when all of the particle orbits would be `irregular' and would tend to fill the available phase space. This would mean that the Jeans theorem is irrelevant if not wrong (cf Binney, 1982).  

We proceed to discuss the special cases of spherical  symmetry in physical space (not necessarily isotropic). Such limits are important since we expect for example the spherical isothermal case to provide the accretion equilibrium for all `flat' cases ($\ad\le 1$) and since spherical spatial symmetry in general has been claimed to minimize the energy of collisionless systems (Aly, 1989) subject to a reasonable phase space volume constraint.  

\subsection{Spherical symmetry with anisotropic velocities\label{sec:sphsym}}

With the assumption of spherical symmetry in real space but not in velocity space,  $A$ is a constant and equation (\ref{res21}) yields immediately;
\be
P_{oo}=B(\kappa_i)(\zeta_\perp^2)^{-q(\ad)},\label{res3}
\ee
where the $\kappa_i$ denote any other integrals. In fact the third, fourth and fifth of equations (\ref{gsphch}) may be combined to yield another integral in the form
\be
\kappa= (\zeta_\perp^2)^{\frac{(\ad-1)}{(2-\ad)}}\left(\frac{\zeta_R^2+\zeta_\perp^2}{2}+\frac{4\pi G A}{i(\ad)}\right).
\ee
This appears to be the only other explicit integral. Using the various definitions we obtain 
\be
\kappa=(j_\perp^2)^{\frac{(\ad-1)}{(2-\ad)}}~E,\label{integral2}
\ee
where $j_\perp^2\equiv (rv_\perp)^2$ and the specific energy is 
\be
E\equiv \frac{v_r^2+v_\perp^2}{2}+\frac{4\pi G A}{i(\ad)}r^{(2-2\ad)}.
\label{specen}
\ee
Consequently we find the steady self-similar DF in the form

\be
f=B(\kappa,\kappa_i)(j_\perp^2)^{-q(\ad)},
\label{DFgen}
\ee
in agreement with Jeans' theorem once again. But we know that in general the DF has the form of equation (\ref{genDF2}) and  we find this to be so if we take 
 $B(\kappa,\kappa_i)$ to be 
\be
B(\kappa,\kappa_i)=B(\kappa_i)\times (|\kappa|)^{\frac{(3-\ad)}{2(\ad-1)}},
\ee
for then the DF above takes the general form  
\be
f=B(\kappa_i)\times |E|^{\frac{(3-\ad)}{2(\ad-1)}},
\label{DFspec}
\ee
where now $\kappa$ itself may be one of the $\kappa_i$. Thus the DF of equation (\ref{DFspec}) is clearly not unique in spherical symmetry. In the case of $\ad=2$, the form is given correctly by equation (\ref{DFspec}), but the additional integral is $\zeta_\perp^2$. The inhibiting effect of angular momentum on the relaxation was already emphasized by Aly (1989). The DF (\ref{DFspec}) was already found in Henriksen and Widrow (1995) as part of a family of strictly steady self-similar solutions.

We remark here that the preceding results are those that would have been found in paper I for this case, had an algebraic error not occurred there.
It is easy to demonstrate this fact by applying our procedure directly to the spherically symmetric form of the CBE as in paper I namely 
\be
\partial_t f + v_r\partial_r f+(\frac{j_\perp^2}{r^3}-\partial_r\Phi)\partial_{v_r}f=0,
\ee
where the tangential angular momentum is $j_\perp^2\equiv r^2v_\perp^2$ and all derivatives are taken while holding it constant. The procedure followed above leads directly to the conclusion of equation (\ref{DFgen}). This corrects the result in paper I and confirms the treatment given here. The result with $\ad=1$ is given in the next section.

In spherical symmetry in an isolated steady state we know that the three components of angular momentum are also isolating integrals, and we might expect these to appear in the ultimate DF, even though only $j_\perp^2$ appears explicitly in our analysis. This is because of the non-isolated nature of the system as it grows which leads to the form (\ref{DFspec}). We note that regularity in $j_\perp^2$ suggests that $\kappa$ does not appear again in equation (\ref{DFspec}), if the function $B$ is restricted to be a power law.

Thus our major conclusion in this section is that  with only spatial spherical symmetry  we can not determine uniquely the DF by looking for stationarity and exact self-similarity alone, unlike the result for the one dimensional systems studied above. Should there be no additional `strange' integrals, and if we require regularity in $j_\perp$ together with at most a scale-free function $B$ in equation (\ref{DFgen}), then we  re-acquire a polytropic form (equation(\ref{DFspec}) with $B$ constant).

\subsection{Spherical symmetry: $\ad=1$ \label{sec:sphsymiso}}

This example is important because it is expected to be the state towards a system ultimately tends while relaxing, and it is not necessarily isotropic. We find on putting $A=0$  equation (\ref{res21}) that 
\be
P_{oo}=B(\kappa)~\zeta_\perp^{-2},\label{Poo}
\ee
where again we find only one additional integral ($ \vec\zeta\equiv \vec v $ here) 
\be
\kappa=\frac{v_r^2+v_\perp^2}{2}-\frac{k}{2}\ln{v_\perp^2},\label{integral4} 
\ee 
and so from the various definitions above with $\ad=1$ we obtain the equilibrium DF as 
\be
f=B(\kappa,\kappa_i)(rv_\perp)^{-2}.\label{sphisoDF}
\ee

We remark that the only choice for $ B(\kappa,\kappa_i)$ that removes a logarithmic singularity in the density (recall that $ d^3v $ may be written $ \pi dv_\perp^2 dv_r$  in spherical symmetry) both as 
\be
 v_\perp^2\rightarrow \infty
\ee 
 and as 
\be
 v_\perp^2 \rightarrow 0
\ee  
is
\be
B(\kappa,\kappa_i)=B(\kappa_i)\times e^{-\frac{2\kappa}{k}}.\label{isochoix}
\ee 
Hence we obtain finally from equation (\ref{sphisoDF}) the well-behaved DF in the Gaussian form
\be
f=B(\kappa_i)\times e^{-\frac{2E}{k}},
\label{Gauss}
\ee
where the specific energy is  
\be
E\equiv \frac{\vec v^2}{2}+k\ln{r}.
\ee
This result now agrees with the general form as found in equation (\ref{DFiso}) but again it is not obviously unique since one of the $\kappa_i$ may be $\kappa$. Since however $\kappa$ is not well defined at the velocity limits, and since we have found no other explicit integrals, we suspect this form to be unique. It is also in agreement with the argument from maximum entropy (Nakamura, 2000).

Assuming this uniqueness, we may replace B in equation (\ref{Gauss}) by a positive constant $b$
and so completely determine this result in physical terms. Thus we re-introduce a fiducial density and a fiducial radius $\rho_o$, $r_o$, so that from the Poisson equation we have $ 4\pi G\rho=k/r^2$ and hence  
\be
k\equiv 4\pi G \rho_o r_o^2.
\ee
Then from the integral of the DF in (\ref{Gauss}) over velocity space (making the usual extention to infinite velocities) we have 
\be
b=(\frac{2}{\pi})^{3/2}\times \frac{1}{4\pi Gr_o^2}\times \frac{1}{\sqrt{k}}.
\ee

Consequently we conclude that in spherical symmetry, the isothermal ($\ad=1$) steady self-similar DF is likely to be a gaussian with the above properties. This is much as in the isotropic case examined in the text. Since theoretical arguments referenced above reveal this DF as a maximum entropy condition, and since we might expect spherical symmetry to be also a condition for equilibrium,  we  suggest this form as a limit towards which collisionless accreting systems tend in many cases. We can not conclude this to be universally so because we have seen that steep initial density profiles (actually $\ad>1$ which is more general) are prevented from reaching this state, apparently because of `trapped' surfaces in phase space. However we note also that it is difficult to distinguish the density profile of a large index polytrope from that of the isothermal sphere as is demonstrated in figure (\ref{fig:poly1}). Our suggestion is that an actual system may be found in a polytropic state with an index (such that $\ad>1$) that increases in time under the continuing excitation due to mergers until it is virtually indistinguishable from a (lowered) gaussian.

\label{lastpage}

\end{document}